\documentclass[aps,prd,floatfix,twocolumn,showpacs,showkeys,groupedaddress,a4paper]{revtex4}  

\RequirePackage[english]{babel}
\usepackage{bm}        
\usepackage{amssymb}   
\usepackage{amsmath}   
\usepackage{graphicx}	
\usepackage{morefloats}
\usepackage{subfigure}
\usepackage{hyperref}

\begin{document}

\title{Unveiling the cosmological QCD phase transition through the eLISA/NGO detector}

\author{V. R. C. Mour\~{a}o Roque}\email{victor.roque@ufabc.edu.br} 

\author{G. Lugones}\email{german.lugones@ufabc.edu.br} \affiliation{Centro de Ci\^encias Naturais e Humanas, Universidade Federal do ABC, Rua Santa Ad\'{e}lia, 166, 09210-170, Santo Andr\'{e}, Brazil.}


\begin{abstract}
We study the evolution of turbulence in the early universe at the QCD epoch using a state-of-the-art equation of state derived from lattice QCD simulations. Since the transition is a crossover we assume that temperature and velocity fluctuations were generated by some event in the previous history of the  Universe  and survive until the QCD epoch due to the extremely large Reynolds number of the primordial fluid. The fluid at the QCD epoch is assumed to be non-viscous, based on the fact that the viscosity per entropy density of the quark gluon plasma obtained from heavy-ion collision experiments at the RHIC and the LHC is extremely small. 
Our hydrodynamic simulations show that the velocity spectrum is very different from the Kolmogorov power law considered in studies of primordial turbulence that focus on first order phase transitions.  This is due to the fact that there is no continuous injection of energy in the system and the viscosity of the fluid is negligible. Thus, as kinetic energy cascades from the larger to the smaller scales, a large amount of kinetic energy is accumulated at the smallest scales due to the lack of dissipation. 
We have obtained the spectrum of the gravitational radiation emitted by the motion of the fluid finding that,
if typical velocity and temperature fluctuations have an amplitude $(\Delta v) /c  \gtrsim 10^{-2}$ and/or $\Delta T/T_c \gtrsim  10^{-3}$, they would be detected by eLISA at frequencies larger than $\sim 10^{-4}$ Hz.  
\end{abstract}

\pacs{98.80.Cq, 98.80.-k, 95.85.Sz, 25.75.Nq, 12.38.Gc, 47.27.-i} 
\maketitle

\section{Introduction}
%
During its evolution the universe passed through several phase transitions. The theory of quantum chromodynamics (QCD) predicts a phase transition in which a hot quark-gluon unconfined phase is converted, as the Universe expands and cools, into a confined hadronic phase. 
The best quantitative evidence for such transition is found in lattice gauge theory of QCD, which shows that this transition occurred at temperatures around $150 - 200$ MeV \cite{Borsanyi2010a,Cheng2010}.
According to recent results, the phase transition for small chemical potentials (condition expected at this epoch for standard cosmology and particle models) is merely an analytic transition (or crossover) \cite{Borsanyi2010}.

Since the QCD transition occurred before the radiation-matter decoupling, the only way to obtain information about it is possibly through gravitational waves created at this epoch. With the eLISA/NGO  (New Gravitational wave Observatory) \cite{Binetruy2012} launch planned for the next years and new detectors being projected, e.g. Big Bang Observatory (BBO) and TOBA \cite{Ishidoshiro2011}, we expect gravitational radiation to be a new source of information about the QCD transition.


Most previous studies on this topic have focused on the effects of a first order transition.  In 1984, Witten \cite{Witten1984} considered the possibility that the QCD phase transition may have produced a detectable gravitational wave signal if the transition is first order and leads to violent bubble collisions. 
Later, Applegate and Hogan \cite{Applegate1985}  made a detailed study of the influence of the QCD transition in the standard cosmological model and studied the relics produced at that period.  Polnarev \cite{Polnarev1985},  demonstrated that polarization and anisotropy in the cosmic background radiation can be induced by gravitational waves depending on their characteristic length (see also \cite{Rotti2012}).
The study of gravitational waves in cosmological phase transitions had a  peak in the early 1990s.  Based on the assumption of a first order transition, several studies were carried out about the nucleation, growth and collision of bubbles  and their relation to the generation of gravitational waves (see e.g. \cite{ Kosowsky1993, Kamionkowski1994, Miller1995}). 
A decade later, works on primordial turbulence continued to be conducted. For example, Kosowsky, Mack and Kahniashvili \cite{Kosowsky2002} calculated the stochastic background of gravitational radiation arising from a period of cosmological turbulence, using a simple model of isotropic Kolmogorov turbulence produced in a cosmological phase transition. Also, Dolgov and Grasso \cite{Dolgov2002} 
showed that an inhomogeneous cosmological lepton number may have produced turbulence in the primordial plasma when neutrinos entered the (almost) free-streaming regime. This effect may be responsible for the origin of cosmic magnetic fields and give rise to a detectable background of gravitational waves.

More recent works have addressed various possibilities for early-universe physics leading to detectable cosmological gravitational wave backgrounds and have analyzed in more detail the turbulent spectrum as well as the spectrum of the gravitational wave signal \cite{Nicolis2004,Kahniashvili2005, Kahniashvili2008,Caprini2006, Kahniashvili2010a,Grojean2007,Gogoberidze2007}. However, to the best of our knowledge, there are no studies of turbulence at the  QCD epoch assuming a crossover transition and employing a realistic lattice QCD equation of state (EoS) for the primordial fluid.  In the present work, we consider an EoS obtained from recent lattice QCD simulations by the Wuppertal-Budapest collaboration  \cite{Borsanyi2010} and perform  hydrodynamic numerical simulations in order to obtain the turbulent spectrum \footnote{One-dimensional-turbulence (ODT) models are widely used as a simulation approach in which high computational costs are mitigated through solution in only a single dimension (see e.g. \cite{Kerstein1999}). This approach has been successfully applied to a variety of problems \cite{odt1}, including for example turbulent diffusion flames \cite{odt2}.} as well as the gravitational radiation generated by the motion of the fluid.  The detectability of the inferred background is discussed in the light of the recently published eLISA/NGO's sensitivity curve \cite{Amaro2012}.


This article is organized as follows: we present firstly the relativistic hydrodynamic equations for a perfect fluid in one dimension  and the here-used EoS (Sec. \ref{sec:hidro}) and we then describe the numerical method employed for the solution of these equations (Sec.\ \ref{sec:metnum}). In Sec.\ \ref{sec:grav}, we present the formalism to obtain the spectrum of gravitational waves from the hydrodynamic evolution of the fluid based on Ref.\ \cite{Weinberg1972}. Finally, we present our results (Sec. \ref{sec:numsim}) and conclusions (Sec. \ref{sec:conclu}).

\section{Basic Equations}
\label{sec:hidro}

\subsection{Hydrodynamics}

To investigate the dynamics and evolution of the primordial fluid, we should solve the equations of hydrodynamics in the context of general relativity. However, as in previous work \cite{Aninos2003} we shall assume a flat space metric filled with a perfect fluid whose stress-energy tensor is defined by
\begin{equation}
\label{eq:TEMfluido}
T^{\mu\nu}=\rho h u^\mu u^\nu + p\eta^{\mu\nu},
\end{equation}
where  $p$ is the pressure, $h$ is the specific enthalpy defined as $h=1+\epsilon+ \frac{p}{\rho}$, $\epsilon$ is the specific internal energy and $\rho$ is the mass density in the rest frame.

In the one-dimensional case adopted here, the hydrodynamic equations written in covariant form consist of three local conservation laws for the above stress-energy tensor  $T^{\mu\nu}$ and the baryon density flow $J^{\mu} = \rho u^{\mu}$:
\begin{equation}
\label{eq:leiscovariante}
 \nabla_\mu T^{\mu\nu} = 0,\quad \nabla_\mu J^{\mu} = 0,
\end{equation}
where we used units in which the speed of light is $c = 1$. The system must be complemented with an equation of state having the functional form  $p = p(\rho,\epsilon)$.

\subsection{Equation of State}

\begin{figure}[t]
\includegraphics[width = .4\textwidth]{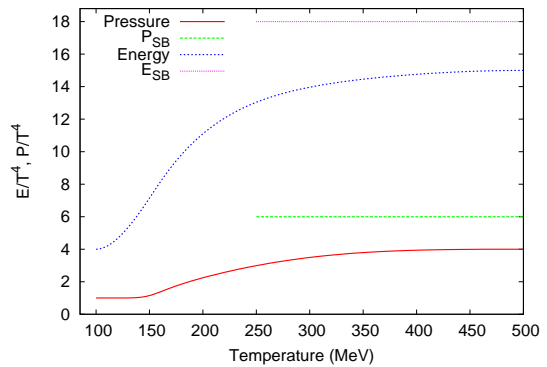}
\caption{EoS describing a crossover obtained using data from lattice QCD \cite{Borsanyi2010} plus the contribution of leptons and photons. The upper dashed line represents the Stefan-Boltzmann limit for a relativistic ideal gas of the same particles  ($p_{SB} \approx 6.77 T^4$ and $E_{SB} \approx 3p_{SB}$).}
\label{fig:EoS}
\end{figure}

In order to construct the EoS for a crossover transition we considered the most recent results of lattice QCD simulations obtained by the Wuppertal-Budapest collaboration (see Ref.\ \cite{Borsanyi2010}), with $N_f = 2 + 1$, i.e. two light (up and down) and one heavy (strange) quarks. To adapt such EoS to the conditions prevailing at the QCD epoch in the early universe we added the contribution of a gas of non-interacting neutrinos ($\nu_\tau, \nu_\mu, \nu_e$), muons, electrons, photons and their antiparticles, all them described by an EoS of the form $E^{\prime} = p^{\prime}/3 = g \pi^2 T^4 / 90$ with $g = 7/8 \times 14 + 2 = 14.24$.  As showed in Ref.\ \cite{Borsanyi2010} the addition of the charm or heavier quarks doesn't introduce any substantial modification in the temperature regime relevant for this work. Such data generate smooth curves for several observables which are typical of a crossover transition. In Fig.\ \ref{fig:EoS} we show the behavior of the pressure $p$ and the energy density $E = \rho (1+\epsilon )$ for a large range of temperatures as well as the Stefan-Boltzmann limit.
%

\section{Numerical Approach}
\label{sec:metnum}
The hydrodynamic equations presented in Eq.\ (\ref{eq:leiscovariante}) can be recast as a hyperbolic system of first-order, flux-conservative equations of the form \cite{Font2002}:
\begin{equation}\label{eq:leisdcons}
 \frac{\partial \boldsymbol{\mathcal{U}}(\mbox{\textbf{w}})}{\partial x^0} + \frac{\partial \boldsymbol{\mathcal{F}^i}(\mbox{\textbf{w}})}{\partial x^i} = 0,
\end{equation}
\noindent where in the laboratory frame the \emph{state vector} $\boldsymbol{\mathcal{U}}$ contains the \emph{conserved} variables $(D, S_i, \tau)$ written in terms of \emph{primitive} variables $\mbox{\textbf{w}} = (\rho, v_i, p)$
\begin{equation} 
\boldsymbol{\mathcal{U}}(\mbox{\textbf{w}}) =  \left[ \begin{array}{l} D \\ S_j\\ \tau \end{array}\right] = 
\left[\begin{array}{c} W\rho \\ \rho h W^2 v_j \\ \rho h W^2 - p - W\rho \end{array}\right],
\end{equation}
\noindent and the \emph{flux vector} $\boldsymbol{\mathcal{F}^i}(\mbox{\textbf{w}})$ is given by
\begin{equation}
 \boldsymbol{\mathcal{F}^i}(\mbox{\textbf{w}}) = \left[ \begin{array}{c} Dv^i\\ S_jv^i + p\delta^i_j\\ \tau v^i + pv^i \end{array} \right],
\end{equation}
\noindent with $W = (1 - v^iv^j)^{-1/2}$ being the Lorentz factor. The elements of the flux matrix are the mass flux $Dv^i$, the momentum flux plus pressure force $S_jv^i + p\delta^i_j$,  and the energy flux plus pressure work $\tau v^i + pv^i$.
Note that the flux vector is written in terms of the conserved variables of the state vector.

The conservation laws written in the form of a Eq.\ (\ref{eq:leisdcons}) allow to solve the problem numerically through a wealth of methods. In the present work we shall employ a \emph{High-Resolution Shock-Capturing} scheme (HRSC)  which is recognized as a very efficient scheme for dealing with complex flows specially when discontinuities are present (see e.g. \cite{Anton2006} and references therein).  These methods use the equations in conservative form together with approximate or exact Riemann solvers to calculate the numerical fluxes between neighboring computational cells. This fact guarantees the capture of all discontinuities (e.g., shock waves) which naturally appear in spatial solutions of the non-linear hyperbolic equations and allows high accuracy in regions where the fluid flow is smooth. These schemes depend on the spectral decomposition of the Jacobian matrix ${\partial \boldsymbol{\mathcal{F}^i}} / {\partial \boldsymbol{\mathcal{U}}}$ which has a complete set of linearly independent eigenvectors (characteristic fields) $\mathbf{r}_i$, and corresponding eigenvalues (characteristic velocities) such that
\begin{equation}
 \left[ \frac{\partial \boldsymbol{\mathcal{F}^i} }{\partial \boldsymbol{\mathcal{U}}}\right] [\mathbf{r}_i] = \lambda_i[\mathbf{r}_i], \qquad  i=0,+,-.
\end{equation}

In the one dimensional case and in a flat Universe the eigenvalues are given by \cite{Font2000,Banyuls1997}
\begin{equation}
\label{eq:valores}
\lambda_0 = v_x,      \qquad     \lambda_{\pm} = \frac{v_x \pm c_s}{1 \pm vc_s},
\end{equation} 
that correspond to the slopes of three families of curves, known as \emph{material}  ($\lambda_0$) and \emph{acoustic}  ($\lambda_\pm$) waves. The relativistic speed of sound $c_s$ for an EoS in which the pressure $p$ is a function of $\rho$ and $\epsilon$ is given by
\begin{equation}
c^2_s = \left.\frac{\partial p}{\partial E}\right|_\mathcal{S} = \frac{\chi}{h}+ \frac{p}{\rho^2}\frac{\kappa}{h} \nonumber
\end{equation}
\noindent where $\chi = \left.\frac{\partial p}{\partial \rho}\right|_\epsilon$, $\kappa = \left.\frac{\partial p}{\partial\epsilon}\right|_\rho$, $\mathcal{S}$ is the entropy per particle, and $E = \rho(1 + \epsilon)$ the total energy density. 

The linearly independent eigenvectors corresponding to each eigenvalue are
\begin{eqnarray} 
\label{eq:vetores_0}
 \mathbf{r}_0 = \left[ \frac{\kappa}{hW(\kappa - \rho c_s^2)},v_x,1-\frac{\kappa}{hW(\kappa - \rho c_s^2)}  \right]^T, \\  
\mathbf{r}_\pm = \left[ 1, hW \lambda_{\pm} \frac{1 - v^2}{1 - v \lambda_\pm} , hW \frac{1 - v^2}{1 - v \lambda_\pm}  - 1 \right]^T.
\label{eq:vetores_pm}
\end{eqnarray}

Our numerical code is based on the Godunov method \cite{Godunov1959} and  the numerical fluxes are obtained using the well known Riemann solver of Roe \cite{Pike1984,Toro2009}: 
\begin{equation}\label{eq:fluxoRoe}
 \mathbf{f}_{i + 1/2} = \frac{1}{2}(\mathbf{f}_r + \mathbf{f}_l) - \frac{1}{2}\sum_{k = 0,\pm} \mathbf{\tilde{r}}_k|\tilde{\lambda}_k| \Delta \tilde{\omega}_k ,
\end{equation}
which has the advantage that can be straightforwardly implemented for an arbitrary EoS. The numerical fluxes $\mathbf{f}_l$ and $\mathbf{f}_r$ are  computed using respectively the primitive variables to the left and right of the $i+\frac{1}{2}$ interface. Both $\tilde{\lambda}_i$ and $\mathbf{\tilde{r}}_i$ are computed by Eqs. (\ref{eq:valores} $-$ \ref{eq:vetores_pm}) using the following weighted quantities:
\begin{eqnarray}
\label{eq:mediasRoe}
 \tilde{\rho} &= &\sqrt{\rho_l\rho_r},\\
\tilde{v}  &= &\frac{\sqrt{\rho_l}v_l + \sqrt{\rho_r}v_r}{\sqrt{\rho_l} + \sqrt{\rho_r}},\\
\tilde{h} &=& \frac{\sqrt{\rho_l}h_l + \sqrt{\rho_r}h_r}{\sqrt{\rho_l} + \sqrt{\rho_r}}.
\end{eqnarray}
The quantities $\{\Delta \tilde{\omega_k}\}$ represent the jumps of the characteristic variables across each characteristic field and are obtained from the inversion of:
\begin{equation}
 \Delta\boldsymbol{ \mathcal{U}} = \boldsymbol{ \mathcal{U}}_r - \boldsymbol{ \mathcal{U}}_l = \sum^{3}_{k =1} \Delta \tilde{\mathbf{\omega}}_k\tilde{\mathbf{r}_k}.
\end{equation}
In order to avoid spurious oscillations we use a \textit{monotone upstream centered scheme for conservation laws} (MUSCL),  with a standard ``minmod'' sloper limit \cite{Font:2002kb} for the reconstruction of the cell centered quantities before the computation of the numerical fluxes. 
The integration in time is performed by using a third order strong-stability-preserving Runge-Kutta scheme with five stages \cite{Gottlieb2008}. 

\section{Gravitational Radiation Spectra}
\label{sec:grav}

To determine the gravitational signal emitted by the motion of the primordial fluid,  we adopt the formalism 
presented in Ref. \cite{Weinberg1972}, which has also been used in Refs.\ \citep[among others]{Kosowsky1993,Kamionkowski1994}. All the information needed to calculate the gravitational wave spectrum is contained in purely spatial components of the stress-energy tensor $T^{\mu\nu}(\boldsymbol{x},t)$ that, in our case, is obtained from the hydrodynamic simulation. Following this treatment, the gravitational energy  radiated per solid angle is given by
\begin{equation}\label{eq:energFourier1}
 \frac{dE}{d\Omega} = 2G\Lambda_{ij,lm}(\boldsymbol{\hat{k}})\int^\infty_0 \omega^2 T^{ij*}(\boldsymbol{k},\omega)T^{lm}(\boldsymbol{k},\omega) d\omega ,
\end{equation}

\noindent where
\begin{multline}\nonumber
\Lambda_{ij,lm}(\boldsymbol{\hat{k}}) = \delta_{il}\delta_{jm} - 2 \boldsymbol{\hat{k}}_j \boldsymbol{\hat{k}}_m\delta_{il} + \mbox{$\frac{1}{2}$} \boldsymbol{\hat{k}}_i \boldsymbol{\hat{k}}_j  \boldsymbol{\hat{k}}_l \boldsymbol{\hat{k}}_m \\
 -\mbox{$\frac{1}{2}$}\delta_{ij}\delta_{lm} + \mbox{$\frac{1}{2}$}\delta_{ij} \boldsymbol{\hat{k}}_l \boldsymbol{\hat{k}}_m + \mbox{$\frac{1}{2}$}\delta_{lm} \boldsymbol{\hat{k}}_i  \boldsymbol{\hat{k}}_j,
\end{multline}

\noindent is know as projection tensor. The stress-energy tensor $T^{ij}(\boldsymbol{k},\omega)$ is obtained by a  fast Fourier transform (FFT)  defined as \cite{Weinberg1972,Kamionkowski1994}
\begin{equation}
  T^{ij}(\boldsymbol{k},\omega) = \frac{1}{2 \pi} \iint  d^3x \ dt \ T^{ij}(\boldsymbol{x},t)e^{-i(\boldsymbol{k} \cdot \boldsymbol{x} - \omega t)}.
\end{equation}

Since our hydrodynamic simulations are performed in one spatial dimension, our problem is axially symmetric about the $z$
axis. Thus,   we can take without loss of generality \cite{Kosowsky1992}
\begin{equation}
\boldsymbol{\hat{k}}_x = \sin\theta,\quad  \boldsymbol{\hat{k}}_y = 0, \quad \boldsymbol{\hat{k}}_z = \cos\theta.
\end{equation}
Therefore, the projector tensor reads:
\begin{subequations}
\label{properties}
\begin{align}
 \Lambda_{ij,lm}  & =  \Lambda_{lm,ij} \, ,        \\
 \Lambda_{ij,lm}\,\delta_{ij}  & =  0 \,   ,             \\
 \Lambda_{ij,lm}\,\boldsymbol{\hat{k}}_i \boldsymbol{\hat{k}}_j & =  0 \,  ,   \\
\Lambda_{ij,lm}\,\delta_{iz}\delta_{jz} \delta_{lz}\delta_{mz}  & =  \Lambda_{zz,zz} =  \tfrac{1}{2} ( 1 - \boldsymbol{\hat{k}}_z^2)^2 = \nonumber \\
       & =  \tfrac{1}{2}\sin^4{\theta}.
\end{align}
\end{subequations}

Finally,  the total energy per unit frequency interval emitted as gravitational radiation is given by (c.f. \cite{Kosowsky1992})
\begin{multline}\label{eq:espectro}
\frac{dE}{d\omega} = G\omega^2\int\left|T^{zz}(\boldsymbol{k},\omega)\sin^2{\theta} \right.  \\
 \left. + T^{xx}(\boldsymbol{k},\omega)\cos^2{\theta} - T^{yy}(\boldsymbol{k},\omega)\right|^2 d\Omega.
\end{multline}
 where $T^{zz}(\boldsymbol{k},\omega)$, $T^{xx}(\boldsymbol{k},\omega)$ and $T^{yy}(\boldsymbol{k},\omega)$ are the Fourier transforms of
\begin{equation}
T^{zz}({\boldsymbol{x}},t) = h\rho v_z^2 - p,
\label{Tzz}
\end{equation}
\begin{equation}
T^{xx}({\boldsymbol{x}},t) = T^{yy}({\boldsymbol{x}},t) = p.
\label{Txxyy}
\end{equation}
Integrating Eq. (\ref{eq:espectro}) over the solid angle, the same expression as in Eq. (23) of Ref. \cite{Kosowsky1993} is obtained \footnote{In the present work we focus on the hydrodynamic evolution of a small portion of the primordial fluid with a size of 100 m. Since we perform 1D simulations, our computational domain can be viewed as a cube where the fluid moves only along the z-axis. Thus, the gravitational emission of the domain is not isotropic but depends on the angle with respect to the z-axis, forming an antenna pattern like the ones presented in \cite{Kosowsky1992}. This angular dependence is an artifact of the 1D simulation, but, fortunately, we are interested in the energy density of the gravitational waves, which depends on the angle-averaged emission of the domain. Anyway, differences are to be expected with respect to 2D and 3D simulations, which will be addressed in future work.}
\begin{equation}
 \frac{dE}{d\omega} = \frac{32\pi}{15} G\omega^2\left|T^{zz}(\boldsymbol{k},\omega) -  T^{xx}(\boldsymbol{k},\omega) \right|^2.
\label{eq:finalGW}
\end{equation}
In order to compare our results with eLISA/NGO, we describe  the spectrum in terms of a characteristic amplitude of the stochastic background defined as \cite{Maggiore1999}
\begin{equation}
h_c(f) \equiv 1.3 \times 10^{-18} [\Omega_{GW}(f)h_0^2]^{1/2} \left( \frac{1 \mbox{Hz}}{f}  \right),
\end{equation}
where $f = \omega/ (2 \pi)$,  $h_0 \equiv H_0 / (100 \ \mathrm{km \ s^{-1} \ Mpc^{-1}} )$ being $H_0$ the Hubble constant, and $\Omega_{GW}$ is the energy density of the gravitational waves,  

\begin{equation}
\Omega_{GW} = \frac{1}{\rho_c}\left(  f \frac{dE}{df}\right),
\end{equation}
being $\rho_c = 3H_0^2/8\pi G$  the critical density. Finally, we have to consider the redshift suffered by the waves in their way to the present Universe \cite{Maggiore1999}:
\begin{eqnarray}
f_0 &=& 8\times 10^{-14} f_* \left( \frac{100}{g_*}\right)^{1/3}  \left( \frac{1 \mbox{GeV}}{T_*}\right)  \mathrm{Hz} \label{eq:freq_today}, \\
\Omega_{GW} &=& 1.67 \times 10^{-5} h_0^{-2}  \left( \frac{100}{g_*}\right)^{1/3} \Omega_{GW*}, 
\end{eqnarray}
where the subscript $0$ corresponds to present values and the subscript  $*$ to the values at the epoch of  the transition.

\section{Turbulence in the crossover QCD transition}

Since we are considering a crossover transition at the QCD epoch, we don't expect large perturbations near the  critical temperature as it would be the case for a first order transition. Thus, we assume that fluctuations present at the QCD epoch were generated by some event in the previous history of the  Universe,  e.g. at the electroweak phase transition that occurred at $t \sim 10^{-12} s - 10^{-10} s$. Hence, one may consider that the size of the larger fluctuations at the QCD era are of the order of the size of the horizon at the electroweak era; i.e. we have $a_{EW}  = a_{QCD} \times t^{1/2}_{EW} \times t^{-1/2}_{QCD} \sim 1 \mathrm{m}$.   
In other words, the injection of energy happens at the electroweak scale ($\sim 1$ m) or at an even smaller scale related to inflation \footnote{According to \cite{Schwarz1997}, Seto and Yokoyama \cite{Seto2003}, and Boyle and Steinhardt \cite{Boyle2005} the spectrum of gravitational waves produced at inflation is modified by a variety of post-inflationary physical effects, in particular, deviations from the standard equation of state ($w =1/3$) during the radiation era.  For example, large drops in the number of relativistic particles during a QCD transition would lead to steps in the spectrum of gravitational waves \cite{Schwarz1997}. The study of such reprocessing of the inflationary background during the QCD epoch is out of the scope of the present paper. However, it is necessary to properly understand and disentangle the all post-inflationary effects \cite{Boyle2005}, like the here studied QCD transition, in order to optimally extract the inflationary information in the stochastic gravitational wave background.}, cosmic strings, etc. \cite{Binetruy2012}. This fluctuations are conjectured to survive until the beginning of the QCD phase transition due to the extremely large Reynolds number of the primordial fluid \cite{large_Re} .

In turbulent fluids, we usually have a large stirring length scale $L$ at which turbulent eddies are injected in the fluid. As said above, we are assuming that large fluctuations occur at (or before) the electroweak era and therefore, the stirring scale is  $L \lesssim 1 m$. As the larger eddies break down to smaller ones, there is a cascade of kinetic energy from large to small scales. This cascade effect will stop at a small damping length scale $l_D$ determined by the viscosity of the fluid. Roughly, the damping scale is determined by a Reynolds number close to 1, $\mathrm{Re} \sim 1$. Thus we can estimate the order of magnitude of the damping scale as $l_D \sim \mathrm{Re} \, \nu / v$, where $\nu$ is the kinematic viscosity of the fluid and $v$ its typical velocity. 

The shear viscosity $\eta = \nu (E + p)$ of matter at the QCD epoch can be derived from heavy-ion collision experiments at the Relativistic Heavy-Ion Collider (RHIC) and the Large Hadron Collider (LHC).  In fact, these experiments show that the quark-gluon plasma at temperatures $T_c < T \lesssim 2 T_c$ behaves as a nearly perfect quark-gluon liquid with a  viscosity per entropy density in the range  $1 < 4 \pi (\eta / s)_{QGP} < 2.5$ \cite{Song2011,Heinz2012}, i.e. approaching the Kovtun-Policastro-Son-Starinets lower bound  $\eta / s \geq 1 / (4 \pi)$ \cite{Kovtun2005}. Since the baryon chemical potential and temperature of the RHIC/LHC fluid are not exceedingly different to what it is expected at the early Universe we may assume that the primordial fluid at the QCD transition behaves as a perfect fluid as well. According to this, we would have $l_D \sim 0.1 \, \mathrm{fm}$ which is 14 orders of magnitude smaller than the size of the cell in our numerical simulations, which is  $\Delta x = 0.02 \, \mathrm{m}$. However, the contribution of leptons and photons to the viscosity may be significant.
In fact, theoretical estimations \cite{Ahonen1999} of the viscosity of the quark-gluon plasma (QGP) in the early Universe give values that are roughly one order of magnitude larger than the experimental values obtained at RHIC and LHC.  The viscosity may be up to three orders of magnitude larger if electrons, muons and photons are included in the QGP, and even eight orders of magnitude larger if neutrinos are considered \cite{Ahonen1999}. In spite of this, the viscosity is still very small and the Reynolds number very large, as generally considered in the literature. Actually, when the most viscous composition is considered, the damping length scale is $l_D < 10^{-7}$ m, which is still orders of magnitude smaller than the size of the cell in our numerical simulations.
Therefore, the use of the ideal hydrodynamic equations presented in Sec. \ref{sec:hidro} is fully justified. 

Up to now, most works studying the QCD transition have considered that the distribution of turbulent kinetic energy density has a stationary spectrum consistent with the Kolmogorov phenomenology \cite{Kamionkowski1994,Kosowsky2002,Megevand2008}, i.e. scaling as $\sim k^{-2/3}$ or variations of this \cite{Dolgov2002,Kahniashvili2010}. This spectrum is very frequent in astrophysical environments, like the interplanetary, interstellar or intergalactic medium, which have a significant viscosity and a constant injection of energy at the large scales (induced e.g. by solar flares, supernovae, etc.). This energy is transferred to small scales through a cascading effect until $\sim l_D$,  where kinetic  energy is dissipated into heat due to the fluid viscosity, resulting a  spectrum with a negative slope. 
In the cosmological case, the assumption of a Kolmogorov-like spectrum may be acceptable when there is a continuous injection of energy, as in the case of a first order phase transition, for which bubble collisions introduce a continuous stirring. However, this stirring is not present in a crossover transition, and therefore the assumption of a Kolmogorov spectrum is unjustified. Moreover, since the viscosity of the primordial fluid is tiny, dissipation at small scales does not occur at the same rate as the energy is accumulated  by cascading. Therefore, a very different turbulent spectrum is to be expected.

\begin{figure}[t]
\includegraphics[angle=-90, width = .35 \textwidth]{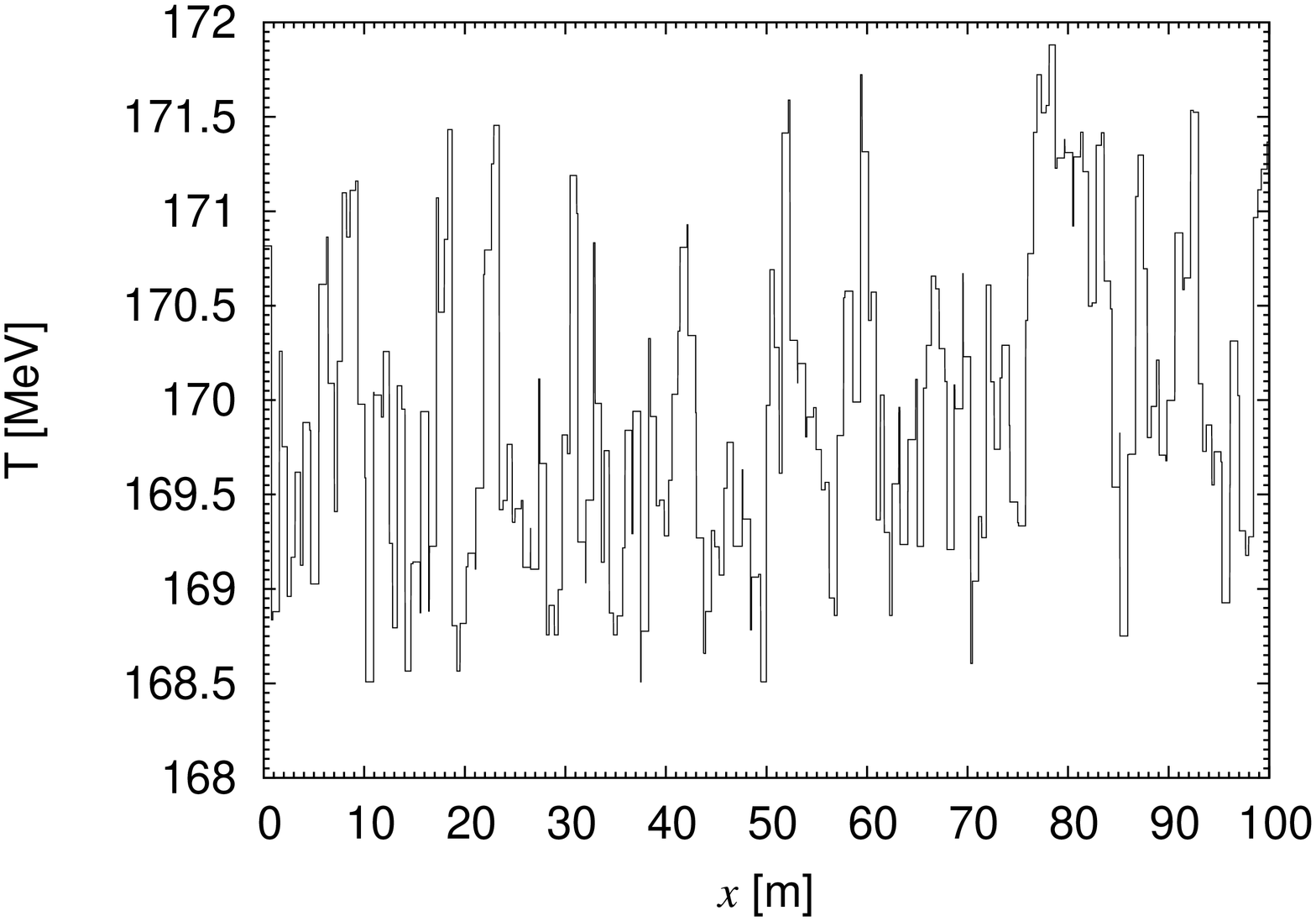}
\includegraphics[angle=-90, width = .35 \textwidth]{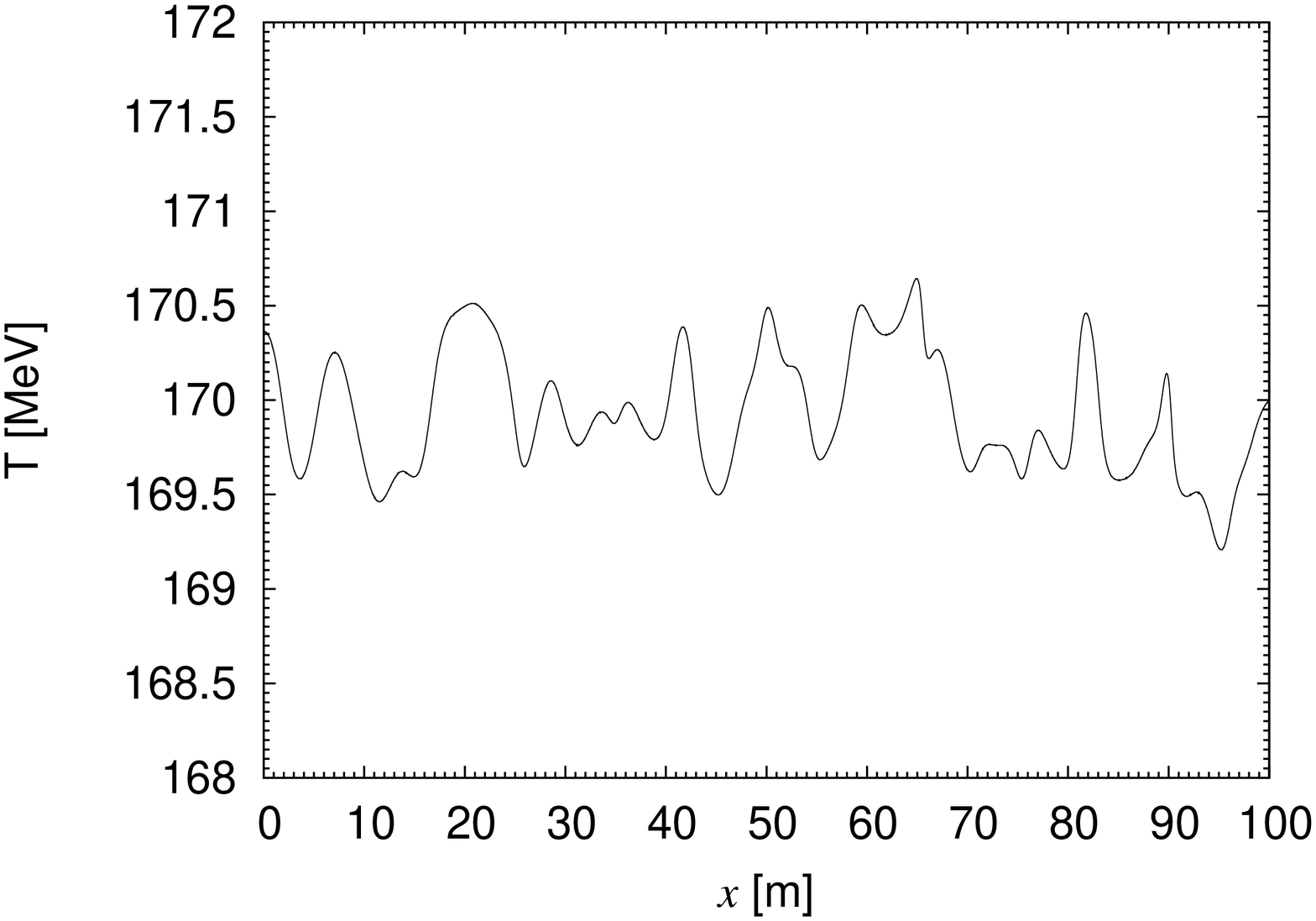}
\includegraphics[angle=-90, width = .35 \textwidth]{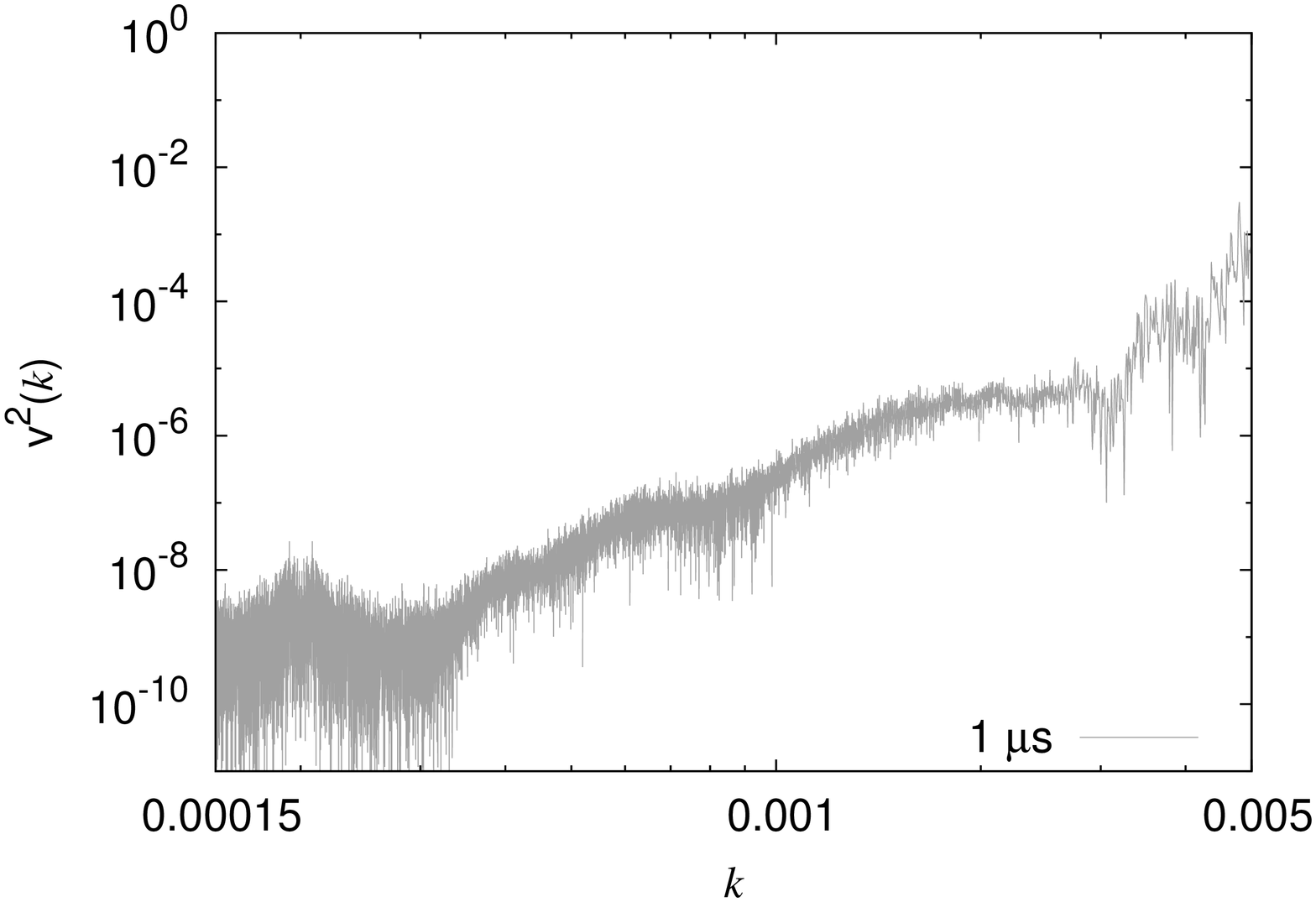}
\caption{\textit{Top}: Initial temperature profile with random temperature fluctuations of maximum amplitude $(\Delta T) /T_c = 10^{-2}$. The fluid is considered at rest at $t=0$.  \textit{Center}: Temperature profile after  $1 \mu s$ of evolution. Temperature inhomogeneities are smoothed and the fluid develops a turbulent motion. 
\textit{Bottom:} Velocity spectrum $ \langle v^2(k) \rangle$ of the turbulent motion of the fluid at $t = 1 \mu s$. There is an energy cascading from the larger to the smaller scales. Energy is not dissipated at the smallest scale because of the negligible viscosity of the primordial fluid. We have considered several different random initial conditions for the temperature profile, all with maximum amplitude  $(\Delta T) /T_c = 10^{-2}$. All of them present the same behavior as presented in this figure.  }
\label{fig:cond_T}
\end{figure}

\section{Results}
\label{sec:numsim}

We have carried out relativistic one-dimensional hydrodynamic simulations for an ideal non-viscous fluid employing the fluid equations and EoS of Sec. \ref{sec:hidro}. We consider a computational domain with a length of  100 m, with 16.384 spatial cells ($2^{14}$) and evolve the system for times larger than $1\,\mu s$. The choice of these intervals is related to the band of the spectrum of gravitational waves that we want to  obtain from the numerical simulations, i.e.  $\sim 10^{-5} - 10^{-3}$ Hz according to the eLISA's sensitivity curve (see below). 
We consider reflective boundary conditions, so that we may see the Universe as a set of various contiguous domains of 100 m. Due to choice of the boundary conditions, disturbances are reflected several times during the simulation, representing the fluid interaction with neighboring regions having similar profiles.
In agreement with the above discussion, turbulence is included only through inhomogeneities in the initial condition. We have considered three different kinds of random initial profiles: (a)  random temperature inhomogeneities in a fluid at rest, (b) random velocity fluctuations within a fluid with an initial uniform temperature, and (c) random temperature and velocity fluctuations. Since the size of possible disturbances at the time of the transition is unknown we consider fluctuations of maximum amplitude $\Delta T/T_c \sim 10^{-2}, 10^{-3}, 10^{-4}$ around $T_c =170$ MeV and/or $\Delta v/ c \sim  10^{-1}, 10^{-2}, 10^{-3}$ in the initial profile. Our goal is to determine the smallest fluctuation amplitude that would be detected by eLISA, considering the motion of the fluid induced by the initial condition as a source of gravitational radiation.
The hydrodynamic simulations provide the tensor $T(\boldsymbol{x}, t)$ from which we obtain $T(\boldsymbol{k},\omega)$ through a FFT. This allows the calculation of the spectrum of the gravitational radiation emitted by the fluid through Eq. (\ref{eq:finalGW}).

\begin{figure}
\includegraphics[angle=-90, width = .42 \textwidth]{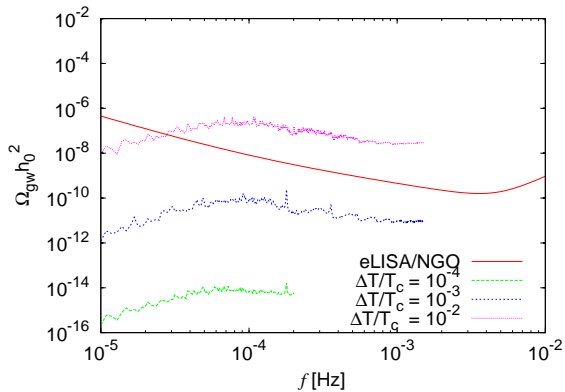}
\caption{Spectra of gravitational waves for hydrodynamic simulations with different initial conditions at the beginning of the QCD epoch. We considered random temperature fluctuations of maximum amplitude $(\Delta T) /T_c =  10^{-2}$ (see Fig. \ref{fig:cond_T}) as well as $10^{-3}$ and $10^{-4}$. For comparison we show the sensitivity curve of eLISA/NGO computed using the expected instrumental noise and the confusion noise generated by unresolved galactic binaries \cite{Amaro2012}.}
\label{fig:randomT}
\end{figure}

In Fig.\ \ref{fig:cond_T} we consider random temperature inhomogeneities with a maximum amplitude of $10^{-2}$ in a fluid that is initially at rest. The initial temperature profile (top panel) presents peaks and valleys of different sizes. As the fluid evolves, these temperature gradients induce the motion of the fluid, temperature inhomogeneities are smoothed (see central panel of Fig.\ \ref{fig:cond_T}), and the fluid develops a turbulent motion with velocities up to $\sim 0.02c$. As discussed before, there is a cascading that transports kinetic energy to the smallest scales where there is no dissipation due to the absence of viscosity. This behavior is apparent in the bottom panel of Fig.\ \ref{fig:cond_T}, which shows that a large amount of kinetic energy is accumulated at the smallest scales (large $k$) at the end of the simulation.  We performed similar simulations with random temperature inhomogeneities with a maximum amplitude of $10^{-3}$ and $10^{-4}$. The hydrodynamic evolution is not shown here because the behavior is qualitatively the same. 

In  Fig. \ref{fig:randomT} we can see a comparison between the gravitational wave spectra arising from the three initial temperature gradients considered here.  In the case with $\Delta T/T_c \lesssim 10^{-2}$ the signal would be  detected by eLISA for a wide range of frequencies. For the simulation with $\Delta T/T_c \lesssim 10^{-3}$ the signal is entirely below but not too far from the eLISA's threshold. In fact, we can show that the signal for fluctuations of about $\Delta T/T_c \lesssim 3 \times 10^{-3}$ is partially overlapping the eLISA's sensitivity curve.  Fluctuations with $\Delta T/T_c \lesssim 10^{-4}$ don't lead to enough motion in the fluid to emit a significant amount of gravitational radiation.

\begin{figure}
\includegraphics[angle=-90, width = .35 \textwidth]{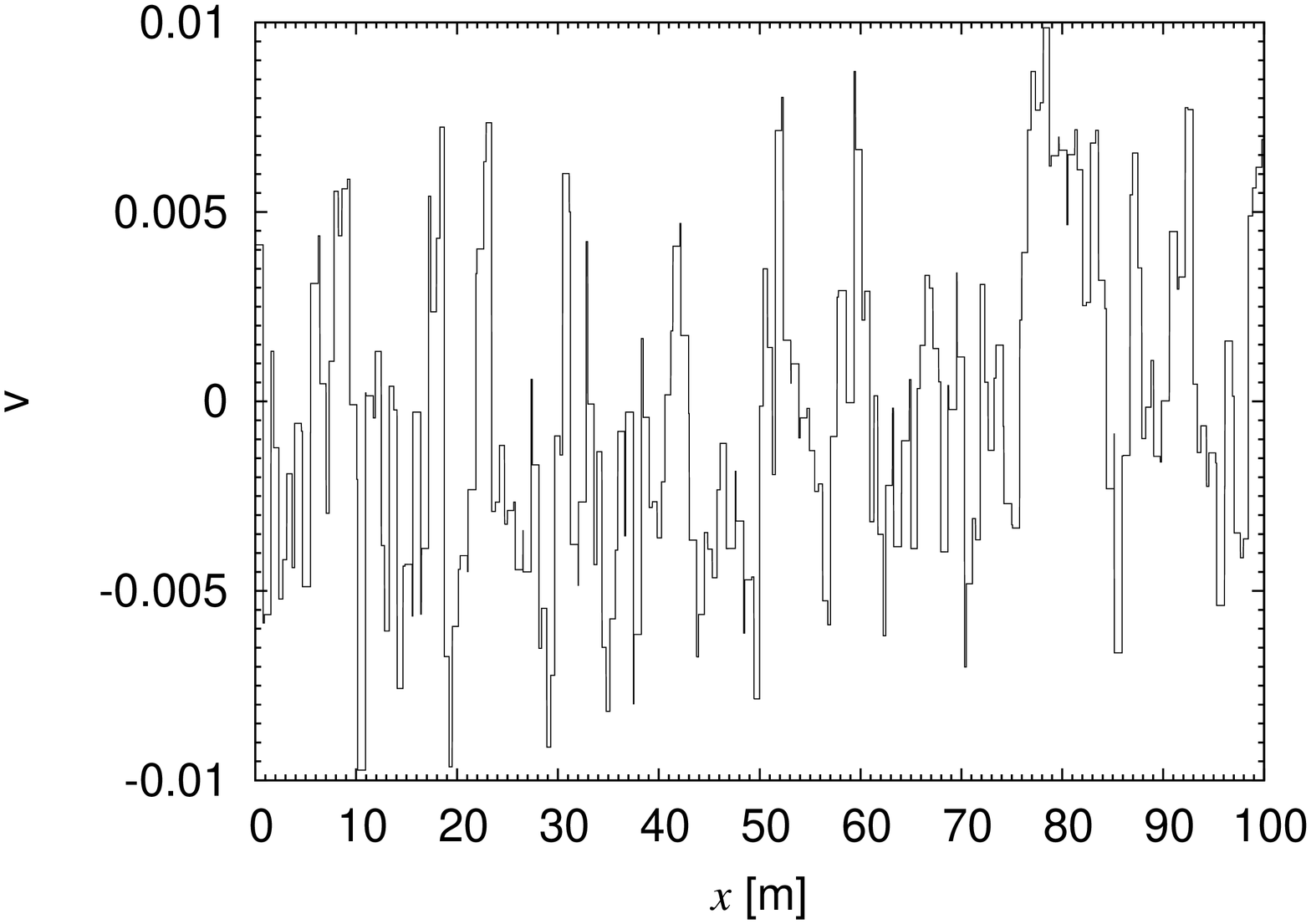}
\includegraphics[angle=-90, width = .35 \textwidth]{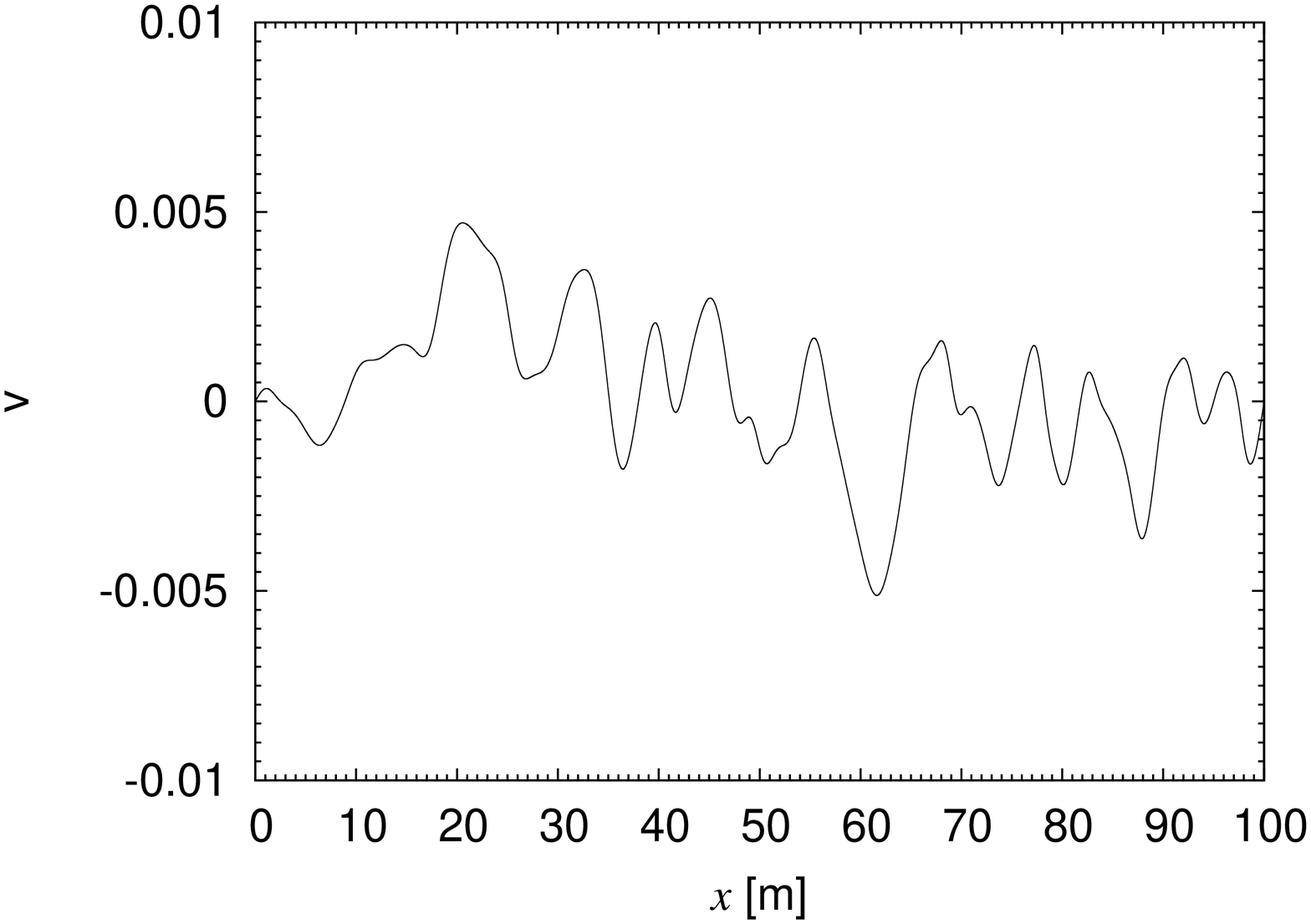}
\includegraphics[angle=-90, width = .35 \textwidth]{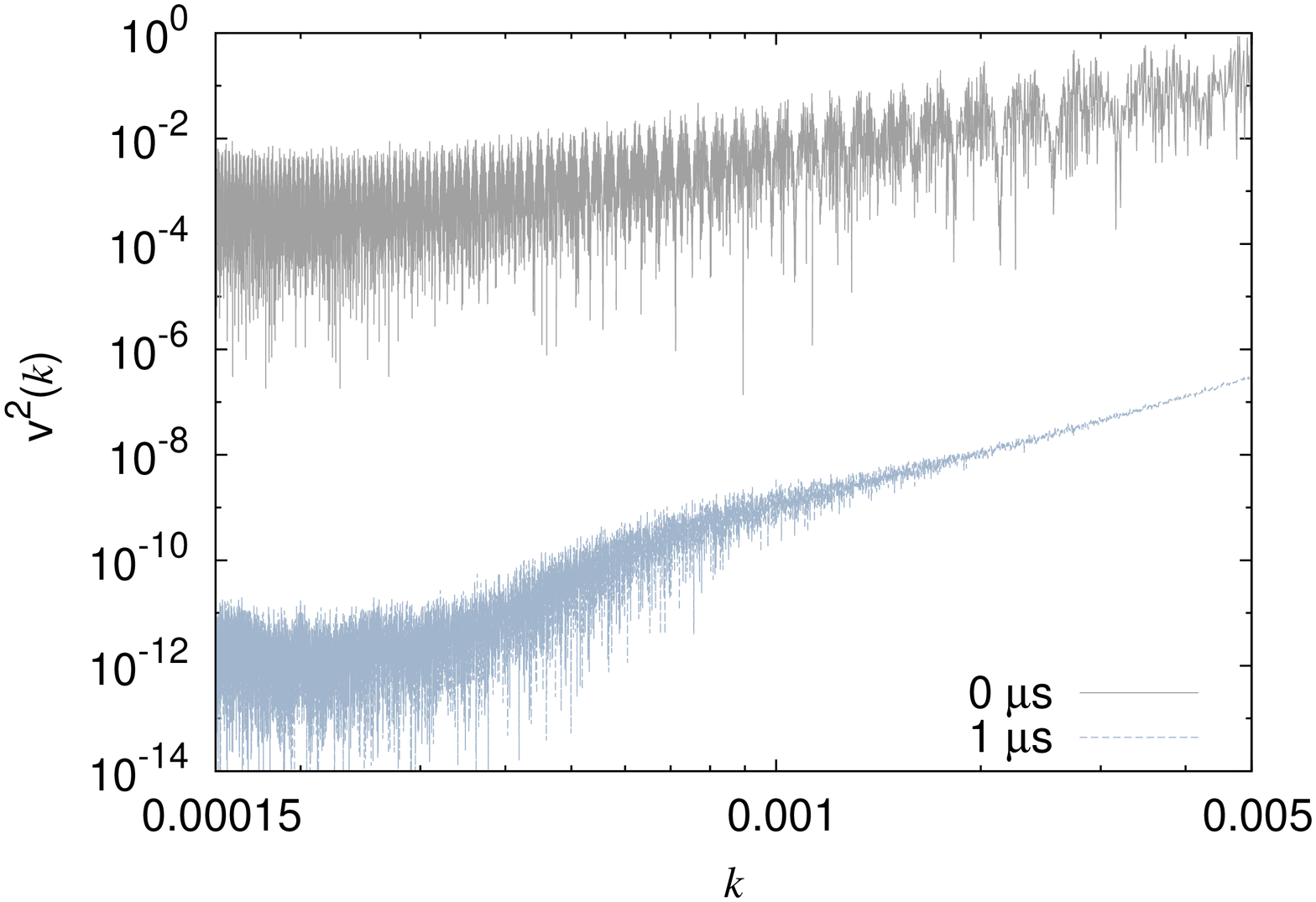}
\caption{\textit{Top}: Initial velocity profile with random velocity fluctuations of maximum amplitude $(\Delta v) /c = 10^{-2} $ and constant temperature $T = T_c \equiv 170$ MeV. \textit{Center}: Velocity profile after  1 $\mu s$ of evolution.  \textit{Bottom:} Velocity spectrum $ \langle v^2(k) \rangle$  at $t = 0$ and $t = 1 \mu s$. The final spectrum is steeper due to the energy cascading from the large to the small scales.}
\label{fig:cond_V}
\end{figure}

\begin{figure}
\includegraphics[angle=-90, width = .38 \textwidth]{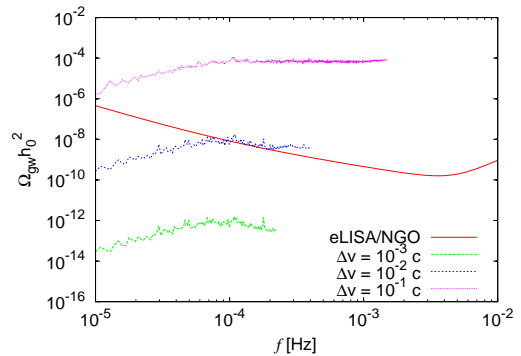}
\caption{Spectrum of gravitational waves for simulations with initial random velocities $(\Delta v) /c = 10^{-3}, 10^{-2}, 10^{-1}$.}
\label{fig:randomV}
\end{figure}

\begin{figure*}
\includegraphics[angle=-90, width = .32 \textwidth]{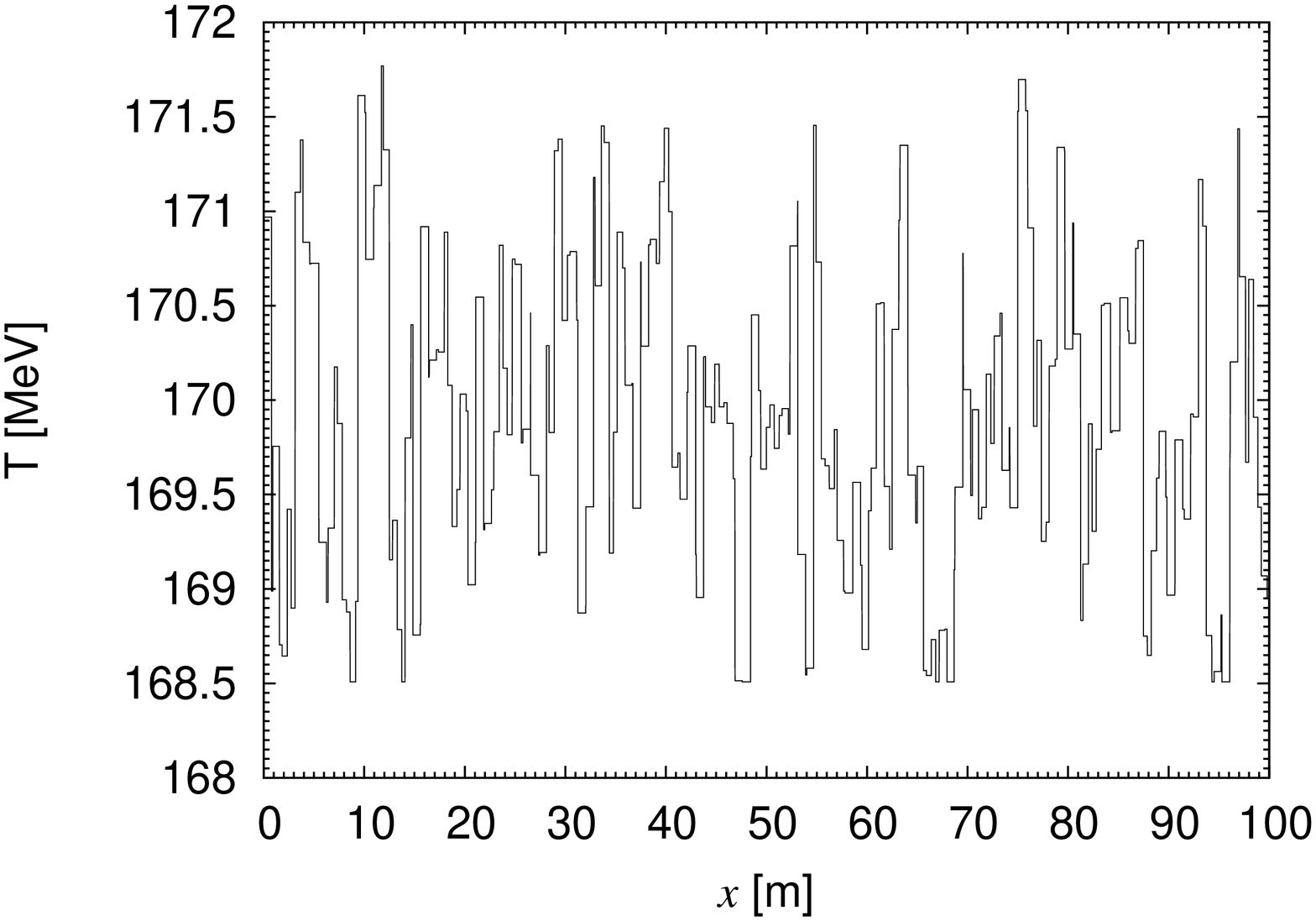}
\includegraphics[angle=-90, width = .32 \textwidth]{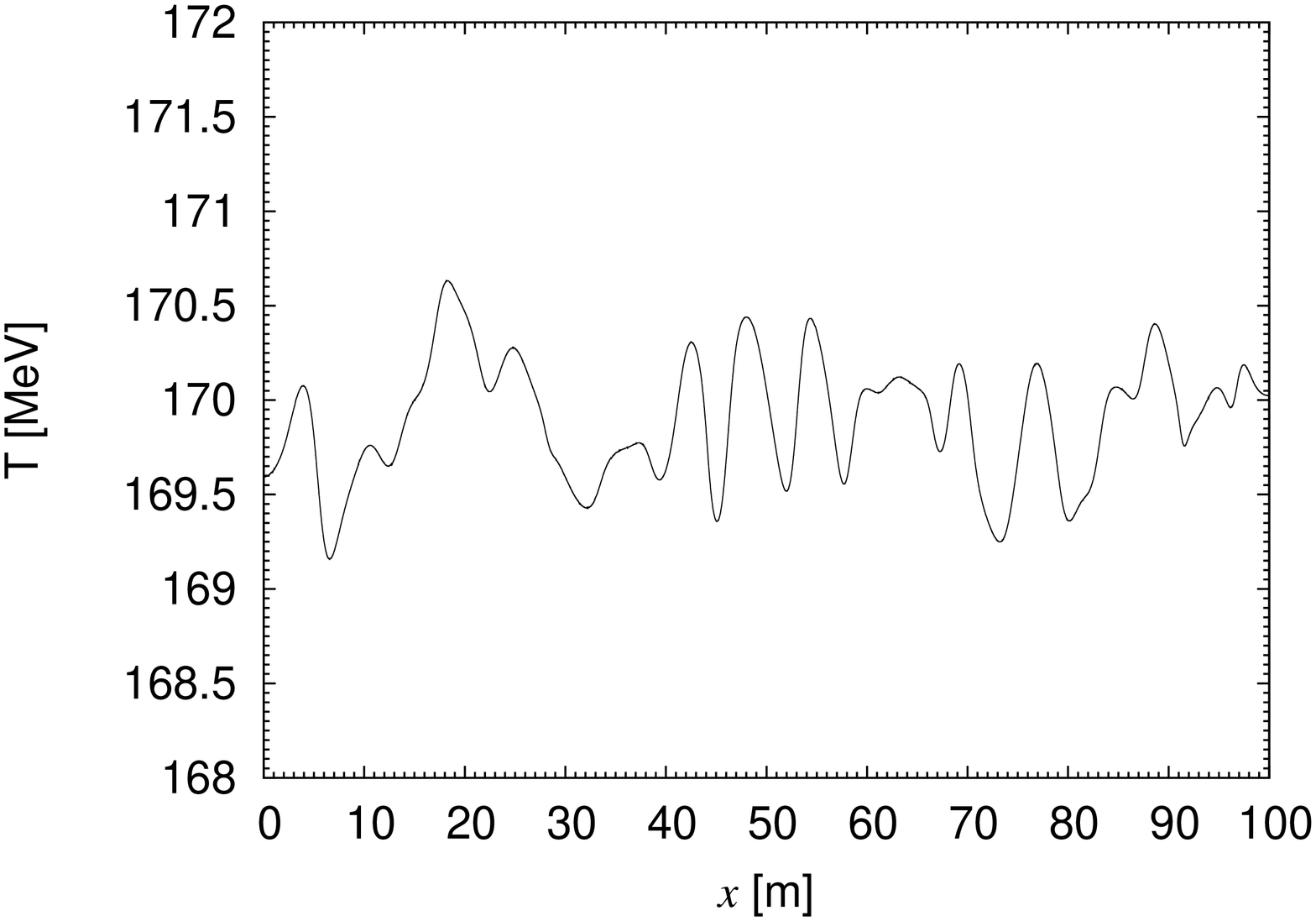}
\includegraphics[angle=-90, width = .32 \textwidth]{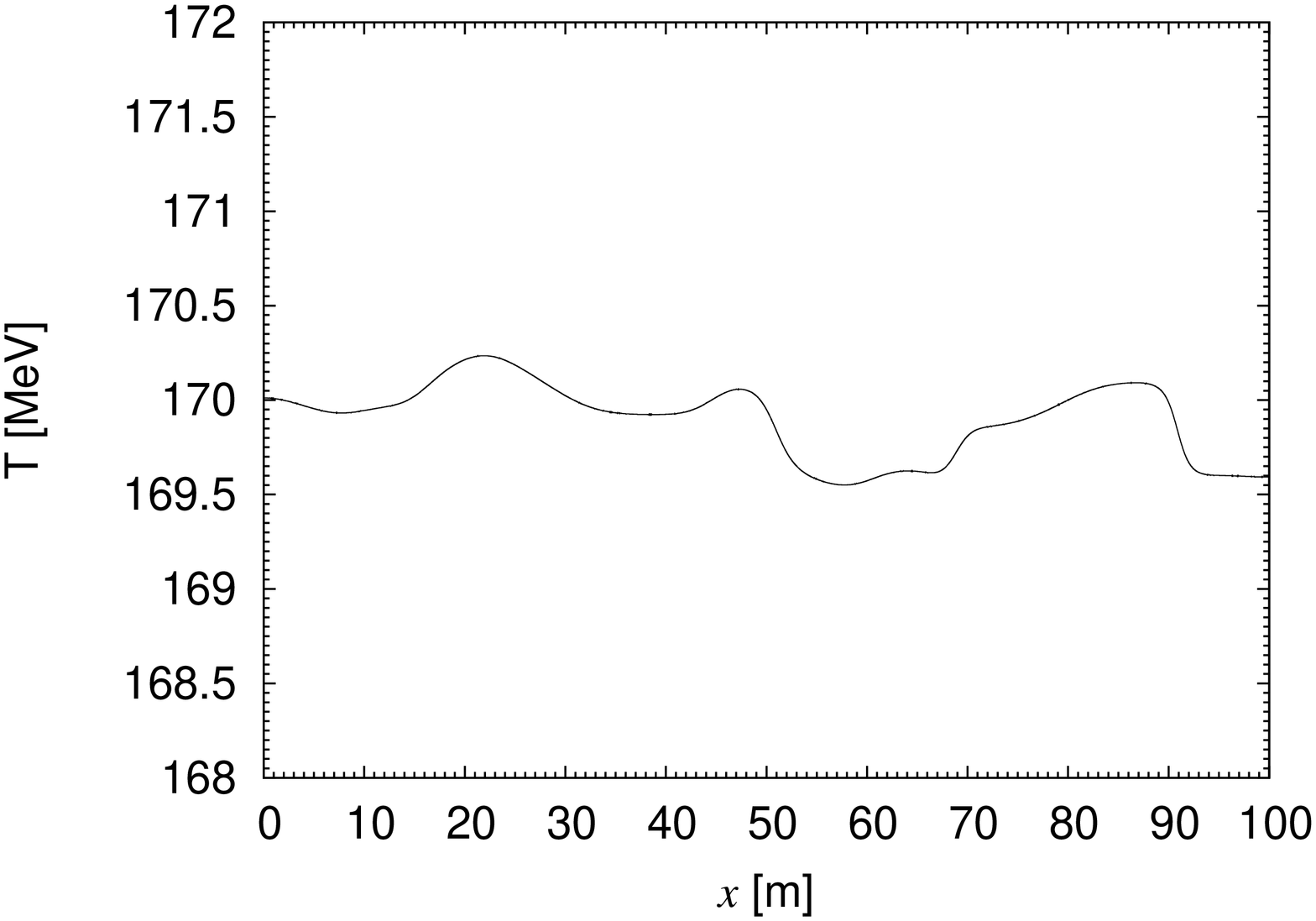}
\includegraphics[angle=-90, width = .32 \textwidth]{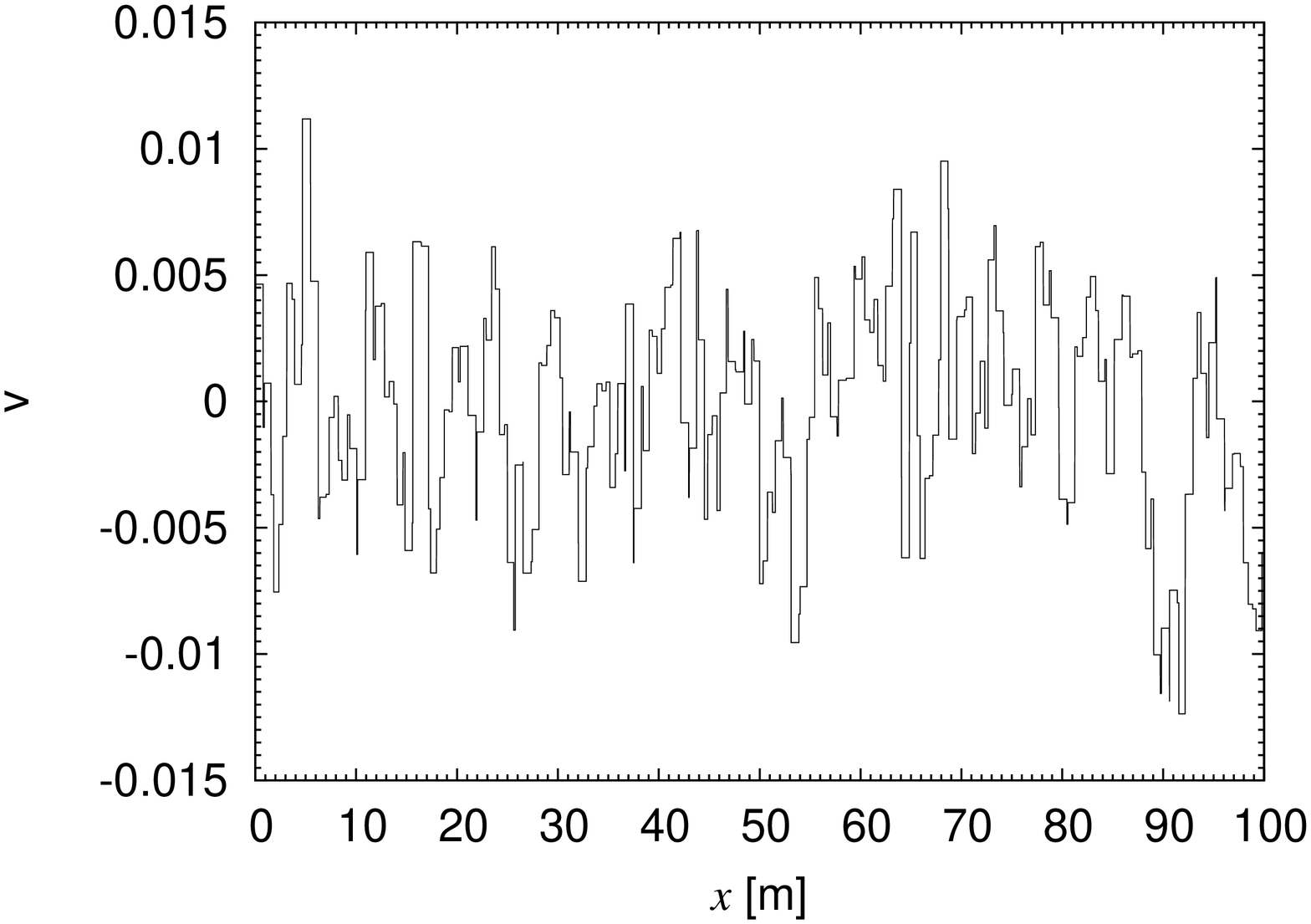}
\includegraphics[angle=-90, width = .32 \textwidth]{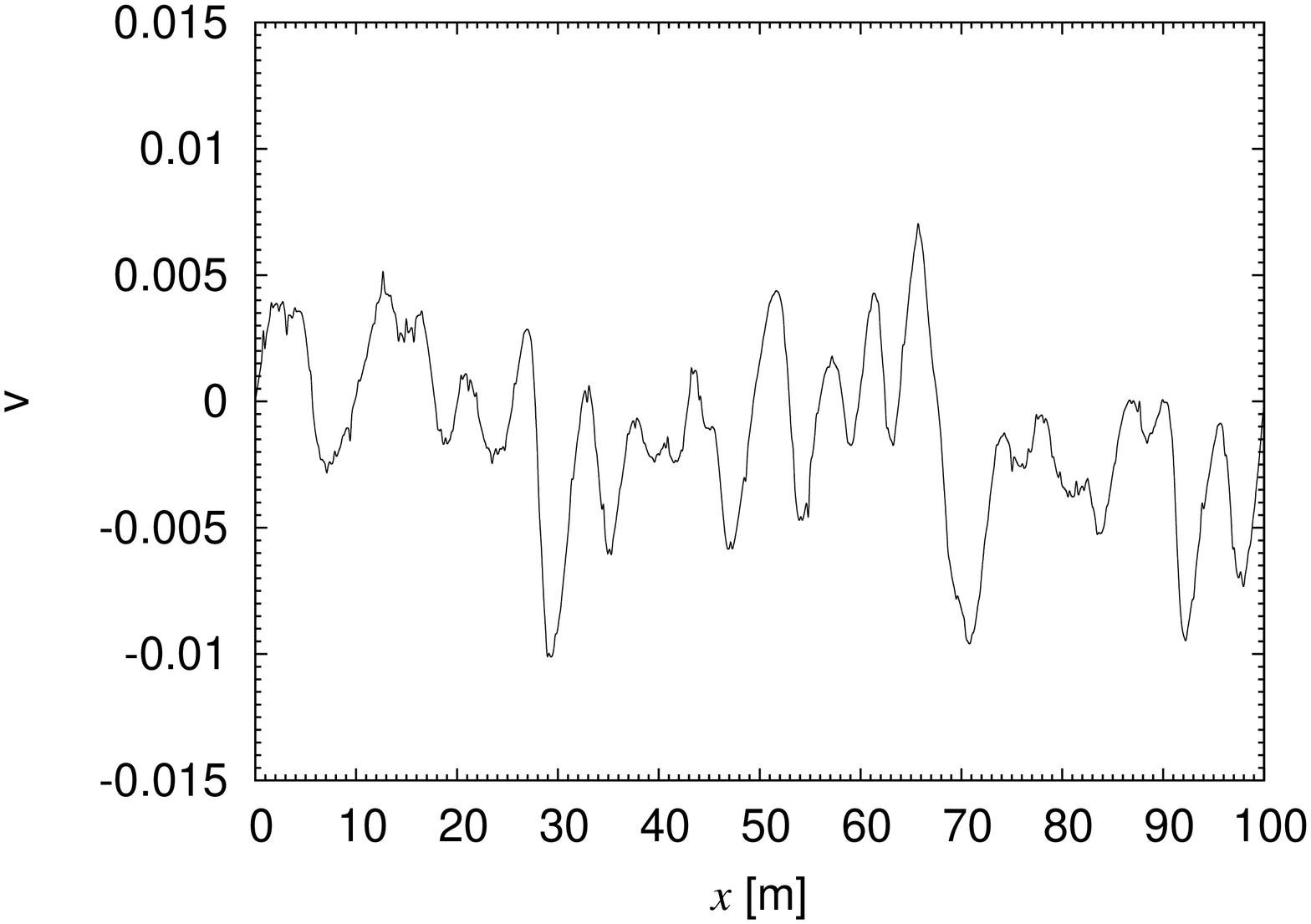}
\includegraphics[angle=-90, width = .32 \textwidth]{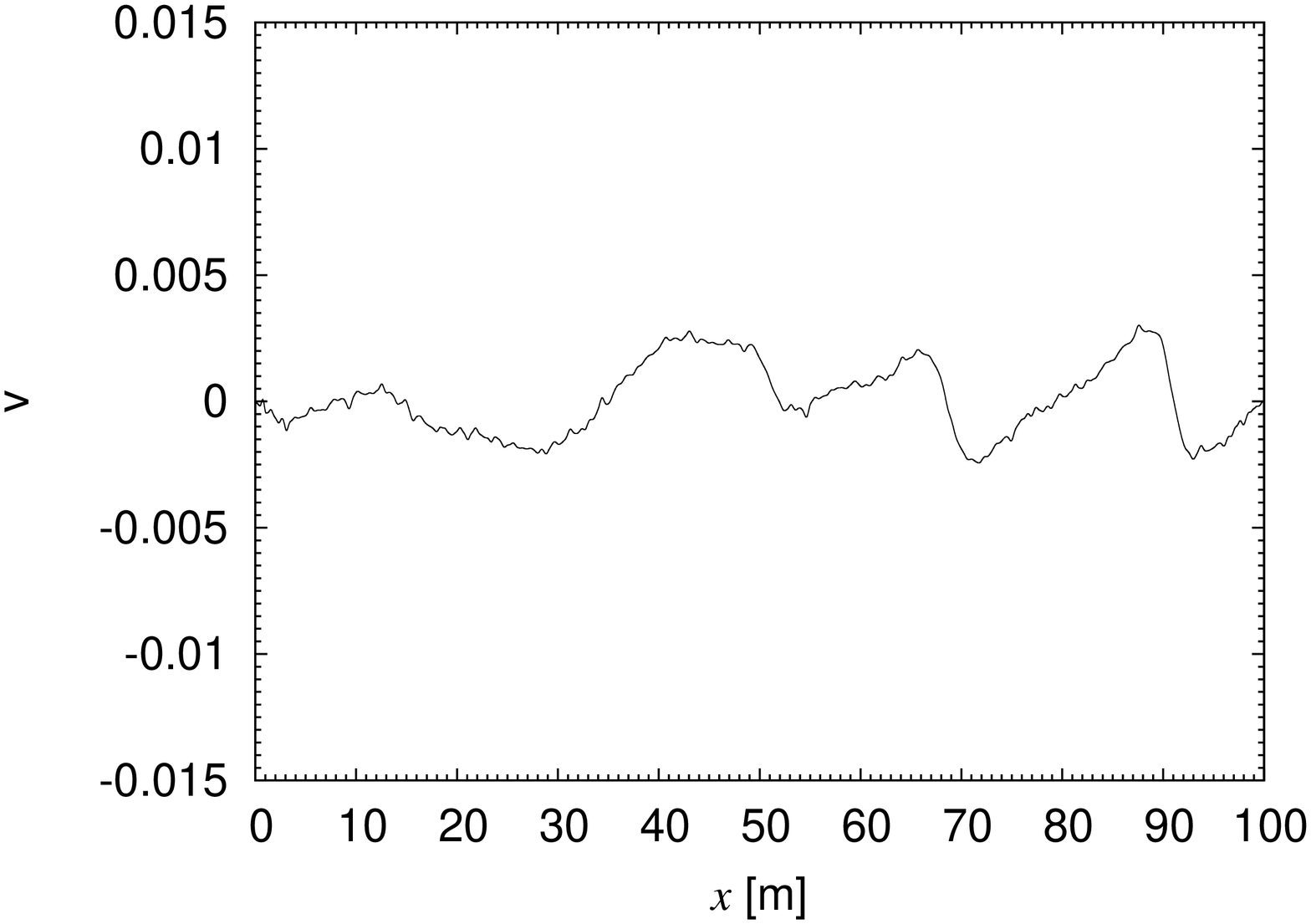}
\caption{\textit{Left panels}:  Initial profiles with random velocity and temperature fluctuations of maximum amplitude $(\Delta v)/c \approx 10^{-2}$ and $(\Delta T) /T_c =  10^{-2}$. \textit{Central panels}: velocity and temperature profiles after  $1 \mu s$ of evolution. \textit{Right panels}: same as before but after  $10 \mu s$ of evolution.}
\label{fig_cond_TV}
\end{figure*}

\begin{figure}
\includegraphics[angle=-90, width = .42 \textwidth]{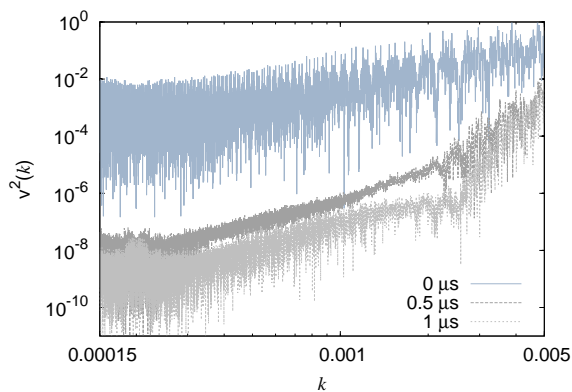}
\caption{Velocity spectrum at different times for the initial conditions of the previous figure.} 
\label{fig:espec_V_TV}
\end{figure}


\begin{figure}
\includegraphics[angle=-90, width = .42 \textwidth]{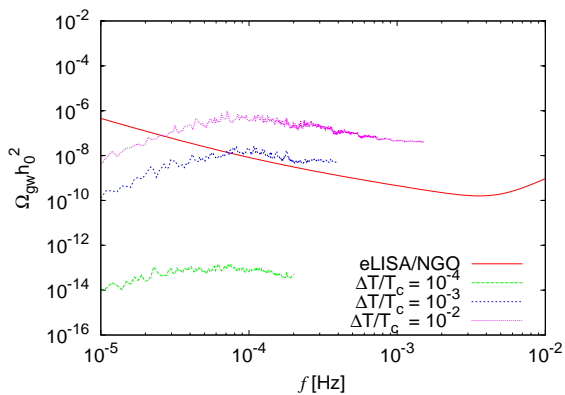}
\caption{Spectrum of gravitational waves for $(\Delta v)/c = 10^{-2}$ and different initial temperature fluctuations.}
\label{fig:randomTV}
\end{figure}

We also considered an initial condition with constant temperature $T= 170$ MeV and a random velocity distribution. 
For an  initial maximum amplitude of $(\Delta v) / c = 0.01$ (see top panel of Fig. \ref{fig:cond_V}) the 
turbulent motion creates temperature gradients of the order of $\Delta T /T_c \approx 3 \times 10^{-3}$ which at the end of simulation  fall to about $\Delta T / T_c \approx 1.5 \times 10^{-3}$ with maximum velocities roughly the half of the initial (see central panel of Fig. \ref{fig:cond_V}).
On the bottom panel of Fig. \ref{fig:cond_V}, we show the initial and final velocity spectra, which present the same behavior as in the previous simulations, confirming the expectation that the spectra must be very different to the Kolmogorov power law. 
For maximum initial velocities of $(\Delta v)/c \sim 0.1$, the induced temperature gradients are an order of magnitude larger than in the previous case and the strong motion of the fluid leads to a large amount of gravitational radiation (Fig.\ \ref{fig:randomV}). 
Notice that the spectrum with $(\Delta v) /c  \lesssim 10^{-2}$ is almost overlapping the eLISA sensitivity curve for frequencies larger that $10^{-4}$ Hz as in the simulation with initial condition $\Delta T/T_c < 3 \times 10^{-2}$, $(\Delta v) /c = 0$. In spite of these initial conditions being very different, the system evolves in both cases to similar temperature and velocity profiles in a timescale considerably smaller than the total time of the simulation. 
In fact, simulations with concomitant velocity and temperature gradients in the initial condition give essentially the same results as before, i.e. a progressive homogenization of both the temperature and velocity profiles, an energy cascade to the smaller scales, and a gravitational wave spectrum with a peak around $10^{-4}$ Hz (see Figs. \ref{fig_cond_TV}$-$\ref{fig:randomTV}). This peak in the spectrum is due to 
an exponential decay of the fluid velocity. Actually, we verified that $|v| \sim \exp(-t/\tau)$ with $\tau \sim 10^{-8}$s for most values of the wave number $k$, showing that most of the fluid motion happens within this time interval. Since gravitational wave emission is associated with the level of turbulent motion in the fluid, we expect a maximum in the spectrum at a frequency  $\sim \tau^{-1}$, which from Eq. (\ref{eq:freq_today}) turns out to be $\sim 10^{-4}$ Hz; i.e. consistent with the maximum in the spectra of Figs. \ref{fig:randomT}, \ref{fig:randomV} and \ref{fig:randomTV}.

\section{Summary and Conclusions}
\label{sec:conclu}

In this paper, we have studied the evolution of primordial turbulence in the universe at the QCD epoch using a state-of-the-art equation of state derived from lattice QCD simulations \cite{Borsanyi2010}.
Since the transition is a crossover, we don't expect large perturbations near the  critical temperature as it would be the case for a first order transition. Thus, we assume that temperature and velocity fluctuations  were generated by some event in the previous history of the  Universe  (e.g. a first order transition at the electroweak scale or at an even smaller scale related to inflation, cosmic strings, etc.) with typical sizes
smaller than the size of the horizon at the electroweak era, $L \lesssim 1 m$. Due to the extremely large Reynolds number of the primordial fluid we consider that these inhomogeneities are able to survive until the QCD epoch.

The primordial fluid at the QCD epoch is assumed to be non-viscous, based on the fact that the viscosity per entropy density of the quark gluon plasma obtained from heavy-ion collision experiments at the RHIC and the LHC is extremely small, more specifically, in the range  $1 < 4 \pi (\eta / s)_{QGP} < 2.5$ at temperatures $T_c < T \lesssim 2 T_c$  \cite{Song2011,Heinz2012}. For the hydrodynamic simulations, we considered a one dimensional computational domain with a length of  100 m divided in $2^{14}$ cells,  injected random temperature and velocity fluctuations in the initial conditions, and evolved the system for times larger than $1\,\mu s$. 
Our results show that the velocity spectrum is very different from the Kolmogorov power law considered in most studies of primordial turbulence that focus on first order transitions.  This is due to the fact that there is no continuous injection of energy in the system and the viscosity of the fluid is negligible. Thus, as kinetic energy cascades from the larger to the smaller scales, a large amount of kinetic energy is accumulated at the smallest scales due to the lack of dissipation (see bottom panels of Figs. \ref{fig:cond_T} and \ref{fig:cond_V}, and Fig. \ref{fig:espec_V_TV}). 
 
We have obtained the spectrum of the gravitational radiation emitted by the motion of the fluid for different initial profiles that include random temperature and velocity fluctuations of different maximum amplitudes.
We find that if typical fluctuations have an amplitude $(\Delta v) /c  \gtrsim 10^{-2}$ and/or $\Delta T/T_c \gtrsim  10^{-3}$, they would be detected by eLISA at frequencies larger than $\sim 10^{-4}$ Hz.

\begin{acknowledgments}
V. R. C. Mour\~ao Roque acknowledges the financial support received from UFABC and CAPES. G. Lugones acknowledges the financial support received from FAPESP.
\end{acknowledgments}

\end{document}